\documentclass[journal,10pt,twoside,web]{ieeecolor}
\usepackage{generic}
\usepackage{cite}
\usepackage{amsmath,amssymb,amsfonts}
\usepackage{algorithmic}
\usepackage{graphicx}
\usepackage{textcomp}
\usepackage{soul}

\usepackage{graphics} 
\usepackage{epsfig} 
\usepackage{mathptmx} 
\usepackage{times} 
\usepackage{breqn}
\usepackage{array}
\usepackage{flushend}
\usepackage{mathtools}
\usepackage{bm}
\usepackage{enumerate}
\usepackage{breqn}
\usepackage{cite}
\usepackage[T1]{fontenc}
\usepackage[ruled,vlined,noresetcount]{algorithm2e}
\usepackage{mathtools}
\usepackage{bbm}
\usepackage{float}
\usepackage{pifont}
\usepackage{microtype}
\usepackage{subfigure}
\usepackage{float}

\newtheorem{thm}{\textit{Theorem}}
\newtheorem{lem}{Lemma}

\newtheorem{rem}{Remark}

\newcommand{\be}{\begin{equation}}
\newcommand{\ee}{\end{equation}}
\newcommand{\ben}{\begin{equation*}}
\newcommand{\een}{\end{equation*}}
\newcommand{\bea}{\begin{eqnarray}}
\newcommand{\eea}{\end{eqnarray}}
\newcommand{\bean}{\begin{eqnarray*}}
\newcommand{\eean}{\end{eqnarray*}}
\newcommand{\X} {{\mathcal X}}
\renewcommand{\H} {{\mathcal H}}

\newcommand{\B} {{\mathcal B}}

\newcommand{\N} {{\mathcal N}}

\newcommand{\R} {{\mathbb R}}
\newcommand{\D} {{\mathbb D}}

\newcommand{\diag} {{\rm diag}}
\newcommand{\K} {{\hat K}}

\renewcommand{\d} {{\delta}}

\def\BibTeX{{\rm B\kern-.05em{\sc i\kern-.025em b}\kern-.08em
    T\kern-.1667em\lower.7ex\hbox{E}\kern-.125emX}}
\begin{document}
\title{Computationally Efficient Robust Model Predictive Control for Uncertain System using Causal State-Feedback Parameterization}
\author{Anastasis Georgiou, \IEEEmembership{Member, IEEE}, Furqan Tahir, Imad M. Jaimoukha and Simos A. Evangelou, \IEEEmembership{Senior Member, IEEE}
\thanks{This work has been funded by an EPSRC Industrial CASE Studentship award in collaboration with Schlumberger (EP/R512540/1).}
\thanks{A. Georgiou, S.A. Evangelou and I.M. Jaimoukha are with the Department of Electrical and Electronic Engineering, Imperial College, London SW7 2AZ, U.K. \tt\small(anastasis.georgiou16@ic.ac.uk, s.evangelou@ic.ac.uk, i.jaimouka@ic.ac.uk). }
\thanks{F. Tahir is with Voltaware Services Limited, London SW5 9AS, UK. {\tt\small (furqan.tahir07@alumni.imperial.ac.uk)}. }
}

\maketitle

\begin{abstract}
This paper investigates the problem of robust model predictive control (RMPC) of linear-time-invariant (LTI) discrete-time systems subject to structured uncertainty and bounded disturbances. Typically, the constrained RMPC problem with state-feedback parameterizations is nonlinear (and nonconvex) with a prohibitively high computational burden for online implementation. To remedy this, a novel approach is proposed to linearize the state-feedback RMPC problem, with minimal conservatism, through the use of semidefinite relaxation techniques. The proposed algorithm computes the state-feedback gain and perturbation online by solving a linear matrix inequality (LMI) optimization that, in comparison to other schemes in the literature is shown to have a substantially reduced computational burden without adversely affecting the tracking performance of the controller. Additionally, an offline strategy that provides initial feasibility on the RMPC problem is presented. The effectiveness of the proposed scheme is demonstrated through numerical examples from the literature.    
\end{abstract}

\section{Introduction}
\label{sec:introduction}
\IEEEPARstart{M}{odel} predictive control (MPC) is a class of receding horizon algorithms in which the current control action is computed by solving online, at each sampling instant, a constrained optimization problem \cite{Rawlings17,Borrelli17, Kouvaritakis16}. MPC is widely used in industry \cite{Qin03} due to its ability to handle multivariable processes and to explicitly consider physical constraints.

Several MPC schemes have been proposed for deterministic systems \cite{Mayne00, Mayne05}. However, uncertainty, in the form of disturbances, state estimation error, plant-model mismatch, and robust constraint satisfaction and stability remain active areas of research \cite{Scokaert98,Kerrigan04,Langson04,Goulart06,Mayne14,Bujarbaruah21}.

The family of MPC algorithms which explicitly take account of uncertainties/disturbances whilst guaranteeing constraint satisfaction and performance is referred to as robust MPC (RMPC). An obvious approach to extend MPC for uncertain systems is to solve an open-loop optimal control problem. Whilst computationally attractive, this may lead to infeasibility and suboptimality~\cite{Mayne00}. A more effective method is to consider a state-feedback control law, as shown in~\cite{Goulart06}, where state-feedback parametrization have been used in RMPC for system subject to additive disturbances. By considering future inputs as linear/nonlinear functions of current and future predicted states, feedback RMPC schemes mitigate the effect of uncertainties whilst potentially avoiding the infeasibility issues. 
The three main types of RMPC schemes include min-max, tube, and LMI based MPC schemes. The min-max MPC method computes the optimal control sequence that satisfies the constraints and steers the system to a robust positively invariant set whilst guarding against worst-case uncertainty~\cite{Scokaert98, Kerrigan04}. The second approach, which has received significant attention, is tube-MPC (TMPC) \cite{Langson04,Kouvaritakis16,Fleming15,RAKOVIC12}. Instead of considering worst-case uncertainty,  these methods first predict a nominal system trajectory and then guarantee that all possible closed-loop state trajectories lie inside a “tube” around the nominal trajectory, where the tube is computed offline using the uncertainty bounds.

The third approach uses semidefinite programming to compute, online, an optimal control  sequence by solving an LMI optimization problem~\cite{Kothare96,Cuzzola02,Tahir11,Tahir12}. The advantages of the LMI based method are the explicit incorporation of uncertainty and the polynomial time required for its solution, which, though still high compared with min-max and TMPC methods, allows online implementation~\cite{Boyd1997} for certain problems. Further details around the implementation issues and trade-offs were reviewed and quantified in~\cite{Gesser18}.

In the work described above, the focus was on systems involving only disturbances or scalar uncertainties. A generalization to systems with structured uncertainties and disturbances was proposed in tube-MPC format in \cite{Carpintero13,RAKOVIC13,HANEMA20} and in ongoing research on System Level Synthesis \cite{Dean19,Chen20}. An LMI based RMPC approach was proposed in \cite{Tahir13} and used in an industrial directional drilling application in ~\cite{Georgiou20}. In this scheme, the state-feedback gain and control perturbation are computed online whilst avoiding the nonconvexity issues. Although this approach shows significant performance improvement, it introduces a large online computational burden, which makes it unsuitable for fast dynamical systems. To solve this problem, an extension to \cite{Tahir13} is proposed here that has a substantially reduced online computational burden without sacrificing performance. 

The contributions of this work are summarized as follows. Firstly, a new LMI-based RMPC scheme is proposed in Section \ref{sec:Problem} for systems subject to structured uncertainty and disturbances. The feedback gain and control perturbation are considered as decision variables whilst nonlinearities are circumvented using a novel linearization procedure (see Section~\ref{sec:Linearization}). This substantially reduces the online computations, while it improves performance due to its less restrictive nature (see Remark~\ref{rem:N_linearization}) without reducing the feasibility region. Secondly, to reduce the online computation time further, an extension is proposed in Section~\ref{sec:SingleLMI} which derives a single LMI sufficient condition for all the constraints. This improves the scalability of the algorithm. Finally, Section~\ref{sec:Feasibility} proposes an offline initialization strategy to guarantee initial feasibility for the control problem.

The paper is organised as follows. In Section \ref{sec:Problem}, the uncertain system is described and the RMPC problem is formulated, while nonlinearities and computational intractability are highlighted. In Section~\ref{sec:Linearization}, an LMI solution based on the proposed linearization procedure is provided, with the disturbance recast as uncertainty. A computationally efficient approach for handling constraints by solving a single LMI is presented in Section~\ref{sec:SingleLMI}. Section~\ref{sec:Feasibility} presents a feasibility analysis and an offline policy to guarantee initial feasibility. Simulation results for case studies taken from the literature are presented in~\ref{sec:Alg_Example}, where the effectiveness of the proposed controller in terms of tracking performance, robustness and computational burden is highlighted and compared with current RMPC methods. Finally, a summary is given in Section~\ref{sec:Conclusion}, along with potential future research directions.
The $\mathbb{R}$ denotes the set of real numbers, $\mathbb{R}^n$ denotes the space of $n$-dimensional real (column) vectors, $\mathbb{R}^{n \times m}$ denotes the space of $n\times m$ real matrices, and $\mathbb{D}^n$ denotes the space of diagonal matrices in $\R^{n\times n}$. For $A \!\in\! \mathbb{R}^{n \times m}$, $A^T$ denotes the transpose of $A$. For $A \!\in\! \mathbb{R}^{n \times n}$, $\mathcal{H}(A)\!:\!=A\!+\!A^T$. For $A\!=\!A^T$
we write $A\!\succeq\!0$ ($A\!\preceq\!0$) if $A$ is positive (negative) semidefinite and $A\!\succ\!0$ ($A\!\prec\!0$) if $A$ is positive (negative) definite. For
$x,y\!\in\!\mathbb{R}^n$, the inequality $x\!<\!y$ (similarly: $\leq$, $>$ and $\geq$)
is interpreted element-wise. $I_q$ denotes the $q \times q$ identity matrix with the subscript omitted when it can be inferred from the context. For matrices
$A_1,\ldots,A_m$, $\diag(A_1,\ldots,A_m)$ denotes a block
diagonal matrix whose $i$-th diagonal block is $A_i$. The symbol $e_i$ denotes the $i$-th column of the identity matrix of appropriate dimension. If ${\pmb{U}}\!\subseteq\!\R^{p\times q}$ is a subspace, then ${\bm\B}{\pmb{U}}\!=\!\{U\!\in\! \pmb{U}: UU^T\preceq\! 1\}$  denotes the unit ball of $\pmb{U}$. The Schur complement argument refers to the result that if $C\!\succ\!0$ then
$\left[\!\!\!\!\begin{array}{cc}A&\!\!B\\\ast&\!\!C\end{array}\!\!\!\!\right]\!\!\succeq \!0$ if and only if  $A\!-\!BC^{-1}B^T\!\succeq\!0$, where $\ast$ denotes a term inferred from symmetry. To deal with uncertainty, we use the following lemma based on  \cite{Ghaoui98} and the use of a Schur complement argument.
\begin{lem} \label{Lemma1} Let $H_{11}\!=\!H_{11}^T,H_{12},H_{21}$, and $H_{22}$ be real matrices. Let $\pmb{\widehat{\Delta}}$ be a linear subspace and define the linear subspace:
\begin{equation}\label{Slack_variable}
\widehat{\Psi}\!=\! \{(S,R,G)\!:S,R\!\succ 0,~\!S\Delta\!=\!\Delta R,~\!\H(\Delta G)\!=\!0~\!\forall\Delta\!\in\!\pmb{\widehat{\Delta}}\}.
\end{equation}
Then $det(I\!-\!H_{22}\Delta) \!\neq\! 0$ and $H_{11}\!+\!\H\!\left(H_{12}\Delta(I\!-\!H_{22}\Delta)^{-1}\!H_{21}\right) \!\succ\! 0$ for every $\Delta \!\in\! {\bm\B}\pmb{\widehat{\Delta}}$ if there exists $(S,R,G)\!\in\!\widehat{\Psi}$ such that:
\begin{equation}\label{Suninequality}
\begin{bmatrix}
H_{11} & H_{21}^T +H_{12}G^T &~~H_{12}S  \\[0.00cm]
\ast& ~~~ R+\H\!\left(H_{22}G^T\right) &~~H_{22}S\\[0.00cm]
\ast&\ast&~~S
\end{bmatrix}
\succ 0.
\end{equation}
\end{lem}
We refer to the S-Procedure. This is used to derive sufficient LMI conditions for the sign definiteness of a quadratic function on a set described by quadratic inequality constraints~\cite{Polik07}.
\section{Problem Statement}\label{sec:Problem}
In this section, the system description including control dynamics, constraints and cost signal are first provided. Then the RMPC control problem is presented through recasting disturbance as uncertainty as shown in \cite{Tahir13}. Lastly, the difficulties to solve this optimization problem are highlighted. 
\subsection{System Description}
The following linear discrete-time system, subject to bounded disturbances and norm-bounded structured uncertainty, is considered (see e.g. \cite{Kothare96}):\vspace{-4mm}
\begin{equation}\label{eq:dt_system_model}
\begin{aligned}
\hspace{-5mm}\left.\begin{array}{c}~\\
\begin{bmatrix}
x_{k+1} \\ q_k \\ f_k \\ z_k
\end{bmatrix}\end{array}\right.\!\!\!\!\!\!\!\!\!&=\!\!\!\!\!\!\!\!\!\!\!\left.\begin{array}{rl}~&\!\!\left.\!\!\!\!\!\!\!\!\!\begin{array}{cccc}\scriptstyle{n}&~~\scriptstyle{n_u}&~~\scriptstyle{n_p}&~~~\scriptstyle{n_w}\end{array}\right.\\
\left.\begin{array}{c}\scriptstyle{n}\\\scriptstyle{n_q}\\\scriptstyle{n_f}\\\scriptstyle{n_z}\end{array}\right.&\!\!\!\!\!\!\!\!\!\!\!\!\!\!\begin{bmatrix}
A     & B_u & B_p & B_w\\
C_q & D_{qu} &0&0 \\
C_f & D_{fu}  & D_{fp} & D_{fw}\\
C_z & D_{zu} & D_{zp} & D_{zw}
\end{bmatrix}\end{array}\right.\!\!\!\!\!\!\!\!\!\!\!\!\!\!\left.\begin{array}{c}~\\
\begin{bmatrix}
x_k\\u_k\\p_k\\w_k
\end{bmatrix}\end{array}\right.\!\!\!\!\!\!\!\!\!,\!\!\!\!\!\!&p_k&\!=\!\Delta_k q_k, \\
\begin{bmatrix}
 q_N \\ f_N \\ z_N
\end{bmatrix}\!\!&=\!\!\begin{bmatrix}
\hat{C}_q & 0\\
\hat{C}_f &  \hat{D}_{fp}\\
\hat{C}_z & \hat{D}_{zp}
\end{bmatrix}\!\!\begin{bmatrix}
x_N\\p_N
\end{bmatrix}\!\!, &p_N&=\Delta_N q_N, \\
\end{aligned}
\end{equation}
where $x_k \!\in \!\mathbb{R}^n,u_k \!\in \!\mathbb{R}^{n_u},w_k\! \in\! \mathbb{R}^{n_w},f_k \!\in\! \mathbb{R}^{n_f},z_k\! \in\! \mathbb{R}^{n_z},p_k\! \in\! \mathbb{R}^{n_p}$ and $q_k \!\in\! \mathbb{R}^{n_q}$ are the state, input, disturbance, constraint, cost, and input and output uncertainty vectors, respectively, with $k\!\in\!\N\!:=\!\{0,1,\ldots,N\!-\!1\}$, where $N$ is the horizon length. It is assumed that the state $x_k$ is measurable. Note that the description includes terminal cost and state constraints to ensure closed-loop stability \cite{Rawlings17}. The symbols in capital letters denote coefficient matrices with the dimensions indicated for ease of reference. Furthermore, $\Delta_k\!\in\! \bm\B \pmb{\Delta}$ where $\pmb{\Delta}\!\subseteq\!\R^{{n_p}\times{n_q}}$ is a subspace that captures the uncertainty structure. Note that we allow uncertainty in all the data including the constraints and the cost signal. Finally, the disturbance $w_k$ is assumed to belong to the set $\mathcal{W}_k \!=\! \{ w_k \!\in\! \mathbb{R}^{n_w} \!\!: -\bar{d}_k\!\leq\! w_k \!\leq\! \bar{d}_k\}$, where $\bar{d}_k\!>\!0$ is given.
\subsection{Algebraic Formulation}
To simplify the presentation, we re-parameterize the disturbance as uncertainty by redefining $\mathcal{W}_k\!:=\!\{\Delta_k^w\bar{d}_k\!\!:\Delta_k^w\!\in\!\bm\B\pmb{\Delta}^w\}$, where $\pmb{\Delta}^w\!=\!\mathbb{D}^{n_w}$ and, 
\begin{equation*}
B_p\!:=\!\!\left[\!\!\!\!\begin{array}{cc}B_p&\!\!\!\!B_w\end{array}\!\!\!\!\right]\!\!\!,C_q\!:=\!\!\left[\!\!\!\!\begin{array}{c}C_q\\0\end{array}\!\!\!\!\right]\!\!\!,D_{qu}\!:=\!\!\left[\!\!\!\!\begin{array}{c}D_{qu}\\0\end{array}\!\!\!\!\right]\!\!\!,
\bar{d}_k\!:=\!\!\left[\!\!\!\!\begin{array}{c}0\\\bar{d}_k\end{array}\!\!\!\!\right]\!\!\!,p_k\!:=\!\!\left[\!\!\!\!\begin{array}{c}p_k\\w_k\end{array}\!\!\!\!\right]\!\!\!,
\end{equation*}
$q_k\!:=\!C_qx_k+D_{qu}u_k+\bar{d}_k$
and $n_p\!:=\!n_p\!+\!n_w,n_q\!:=\!n_q\!+\!n_w.$ 

The vector $\bar{z}_k$ is assumed to be given and defines the reference trajectory. The constraint and terminal constraint signals are defined by $\bar{f}_k$ and $\bar{f}_N$, respectively, and are assumed to be known. They are chosen to satisfy polytopic constraints on the input and state  signals, and terminal state signals, respectively. 
The only assumption that is imposed here is that the terminal constraints defined by $\bar{f}_N$ are within a polytopic control invariant set \cite{Chengyuan19}, which can be used to derive condition for recursive feasibility on the control scheme, see Remark~\ref{rec_feas}. 
By defining the stacked vectors,
\begin{equation*}
    \mathbf{u}=\!\!\left[\!\!\begin{array}{c}u_0\\\vdots\\u_{N-1}\end{array}\!\!\right]\!\!\in \mathbb{R}^{N_u},~\mathbf{x}=\!\!\left[\!\!\begin{array}{c}x_1\\\vdots\\x_N\end{array}\!\!\right]\!\!\in \mathbb{R}^{N_n},~\pmb{\zeta}=\!\!\left[\!\!\begin{array}{c}\zeta_0\\\vdots\\\zeta_N\end{array}\!\!\right]\!\!\in \mathbb{R}^{N_\zeta},
\end{equation*}
where $\pmb{\zeta}$ stands for $\mathbf{f},\bar{\mathbf{f}},\mathbf{p},\mathbf{q},\mathbf{z},\bar{\mathbf{z}}$ or $\bar{\mathbf{d}}$ and $N_n=\!Nn,~\!N_u\!=\!Nn_u$ and $N_\zeta\!=\!(N\!+\!1)n_\zeta$, we get
\begin{equation}
\label{Stacked}
\left.\begin{array}{c}~\\
\begin{bmatrix}
\mathbf{x}\\ \mathbf{q} \\ \mathbf{f} \\ \mathbf{z}
\end{bmatrix}\end{array}\right.\!\!\!\!\!\!\!\!\!=\!\!\!\!\!\!\!\!\!\!\!\left.\begin{array}{rl}~&\!\!\left.\!\!\!\!\!\!\!\!\!\begin{array}{cccc}\scriptstyle{n}&~~\!\scriptstyle{N_u}&~~\scriptstyle{N_p}&~~\!\scriptstyle{1}\end{array}\right.\\
\left.\begin{array}{c}\scriptstyle{N_n}\\\scriptstyle{N_q}\\\scriptstyle{N_f}\\\scriptstyle{N_z}\end{array}\right.&\!\!\!\!\!\!\!\!\!\!\!\!\!\!
\begin{bmatrix}
\mathbf{A} & \mathbf{B}_u & \mathbf{B}_p & 0 \\
\mathbf{C}_q  &\mathbf{D} _{qu} & \mathbf{D} _{qp} & \bar{\mathbf{d}} \\
\mathbf{C}_f  &\mathbf{D} _{fu} & \mathbf{D} _{fp} & 0 \\
\mathbf{C}_z  &\mathbf{D} _{zu} & \mathbf{D} _{zp} & 0 \\
\end{bmatrix}\end{array}\right.\!\!\!\!\!\!\!\!\!\!\!\!\!\!\left.\begin{array}{c}~\\
\begin{bmatrix}
x_0 \\ \mathbf{u} \\ \mathbf{p} \\ 1
\end{bmatrix}\end{array}\right.\!\!\!\!\!\!,\quad \mathbf{p} = \hat\Delta \mathbf{q},
\end{equation}
with $\hat\Delta \in \bm\B \pmb{\hat{\Delta}}\subset\R^{N_p\times N_q} $ where,
\bean
    \pmb{\hat\Delta}\!=\!\{{\rm diag}(\Delta_0,\Delta_0^w,\ldots,\Delta_{N\!-\!1},\Delta_{N\!-\!1}^w,\Delta_N)\!\!:\!\Delta_k\!\in\!\pmb{\Delta},\Delta_k^w\!\in\!\pmb{\Delta}^w \}\!,
\eean
and where the stacked matrices in \eqref{Stacked} (shown in bold) have the indicated dimensions and are readily obtained from iterating the dynamics in \eqref{eq:dt_system_model} and the re-definitions in this section.
The input signal $u_i$ is considered as a causal state feedback that depends only on states $x_0,\ldots,x_i$ (see e.g. \cite{Skaf10}). Thus
\begin{equation}
\label{u1}
\mathbf{u}=K_0 x_0+K\mathbf{x}+\pmb{\upsilon},
\end{equation}
where $\pmb{\upsilon} \!\in \!\mathbb{R}^{N_u}$ is the (stacked) control perturbation vector and
$K_0,~\!K$ are the current and predicted future state feedback gains. Causality is preserved by restricting $[ K_0 ~\! K] \!\in\!\mathcal{K}\!\subset\! \mathbb{R}^{N_u \times N_n}$, where $\mathcal{K}$ is the set of $N_u \times N_n$ lower block triangular matrices with $n_u\times n$ blocks. $K_0,K$ and $\pmb{\upsilon}$ are considered as decision variables. Note that, while $K_0$ is redundant for a given $x_0$ as it can be absorbed in $\pmb{\upsilon}$, we keep it for when we tackle the case of variable $x_0$ in Section~\ref{sec:Feasibility}. Substituting the expression of $\mathbf{x}$ in \eqref{Stacked} into \eqref{u1} gives,
\begin{equation}\label{eq:u2}
\mathbf{u}\!=\!\hat{K}_0x_0\!+\!\hat{K}\mathbf{B}_p\mathbf{p}\!+\!\hat{\upsilon},
\end{equation} 
where $\begin{bmatrix}\hat{K}_0 & \!\!\hat{K} & \!\!\hat{\upsilon}
\end{bmatrix}\!\!=\! (I\!-\!K\mathbf{B}_u)^{-1}
\!\begin{bmatrix}
K_0\!+K\mathbf{A} & \!\!K & \!\!\pmb{\upsilon}
\end{bmatrix}\!$. Note that $(I\!-\!K\mathbf{B}_u)$ is invertible due to the lower-triangular structure and that $\mathbf{u}$ is affine in $\hat{K}_0,\hat{K}$ and $\hat{\upsilon}$ which have the same structure as ${K_0,K}$ and $\pmb{\upsilon}$. A standard feedback re-parameterization gives
\begin{equation}\label{eq:Kv}
\begin{bmatrix}
 K_0 & K & \pmb{\upsilon}
\end{bmatrix}= (I+\hat{K}\mathbf{B}_u)^{-1}
\begin{bmatrix}
\hat{K}_0\!-\!\hat{K}\mathbf{A} & \hat{K} & \hat{\upsilon}
\end{bmatrix}\!\!,
\end{equation} 
and so $\left[\hat{K}_0~\hat{K}~\hat{\upsilon}\right]$ will be used as the decision variables instead.
Using (\ref{eq:u2}) to eliminate $\mathbf{u}$ from (\ref{Stacked}) and re-arranging $x_0$ gives
\begin{equation}\label{def1}
\begin{split}
\hspace{-3mm}\left[\!\!\!\begin{array}{c}\mathbf{q}\\\mathbf{f}\\\mathbf{z}-\bar{\mathbf{z}}\end{array}\!\!\!\right]\!\!\!&=\!\!\!\left[\!\!\!\begin{array}{c|c}
\mathbf{D}_{qp}^{\hat{K}}\!\!\!&\!\!\!\mathbf{D}_{q}^{\hat{K}_0,\hat{\upsilon}}\\ \hline
\mathbf{D}_{fp}^{\hat{K}}\!\!\!&\!\!\!\mathbf{D}_{f}^{\hat{K}_0,\hat{\upsilon}}\\ \hline
\mathbf{D}_{zp}^{\hat{K}}\!\!\!&\!\!\!\mathbf{D}_{z}^{\hat{K}_0,\hat{\upsilon}}\end{array}\!\!\!\right]\!\!\!
\left[\!\!\!\begin{array}{c}\mathbf{p}\\1\end{array}\!\!\right]\!\!, \\
&:=\!\!\!
\left[\!\!\!\begin{array}{c|c}
\mathbf{D}_{qp} \!\!+\!\! \mathbf{D}_{qu}\hat{K}\mathbf{B}_{p} \!\!\!&\!\!\! \mathbf{D}_{qu}\hat{\upsilon}\!\!+\!\! (\mathbf{C}_{q}\!\!+\!\!\mathbf{D}_{qu}\hat{K}_0)x_0\!\!+\!\! \bar{d}\\ \hline
\mathbf{D}_{fp}\!\!+\!\!\mathbf{D}_{fu}\hat{K}\mathbf{B}_{p}\!\!\! &\!\!\! \mathbf{D}_{fu}\hat{\upsilon} \!\!+\!\! (\mathbf{C}_{f}\!\!+\!\!\mathbf{D}_{fu}\hat{K}_0)x_0\\ \hline
\mathbf{D}_{zp}\!\!+\!\!\mathbf{D}_{zu}\hat{K}\mathbf{B}_{p} \!\!\!&\!\!\! \mathbf{D}_{zu}\hat{\upsilon}\!\!+\!\! (\mathbf{C}_{z}\!\!+\!\!\mathbf{D}_{zu}\hat{K}_0)x_0\!\!-\!\!\bar{z}\end{array}\!\!\!\right]\!\!\!\!
\left[\!\!\!\!\begin{array}{c}\mathbf{p}\\1\end{array}\!\!\!\right]\!\!\cdot
\end{split}
\end{equation}
Note that all the coefficient matrices in \eqref{def1} are affine in $\hat{K}_0,~\hat{K}$ and $\hat{\upsilon}$. Finally, eliminating $\mathbf{p}$ using $\mathbf{p}=\hat\Delta \mathbf{q}$ we get
\begin{equation}\label{def2}
    \left[\!\!\begin{array}{c}\mathbf{f}\\\mathbf{z}-\bar{\mathbf{z}}\end{array}\right]\!\!=\!\!\left[\!\!\begin{array}{c}
    \mathbf{D}_{f}^{\hat{K}_0,\hat{\upsilon}}+\mathbf{D}_{fp}^{\hat{K}}\hat\Delta(I-\mathbf{D}_{qp}^{\hat{K}}\hat\Delta)^{-1} \mathbf{D}_{q}^{\hat{K}_0,\hat{\upsilon}}\\
        \mathbf{D}_{z}^{\hat{K}_0,\hat{\upsilon}}+\mathbf{D}_{zp}^{\hat{K}}\hat\Delta(I-\mathbf{D}_{qp}^{\hat{K}}\hat\Delta)^{-1} \mathbf{D}_{q}^{\hat{K}_0,\hat{\upsilon}}
    \end{array}\!\!\right]\!\!\cdot
\end{equation}
For convenience, we write $\mathbf{f}\!=\!\mathcal{F}(\hat{K}_0,\hat{K},\hat{\upsilon},\hat\Delta)$ and $(\mathbf{z}\!-\!\bar{\mathbf{z}})^T\!(\mathbf{z}\!-\!\bar{\mathbf{z}})\!=\!\mathcal{Z}(\hat{K}_0,\hat{K},\hat{\upsilon},\hat\Delta)$ to emphasize dependence on the variables.
\subsection{RMPC Problem}\label{RMPC_problem}
Given the initial state $x_0$, the RMPC problem is then to find a feedback law $u_k$ for all $k\in\N$ such that the cost function, is minimized, while the constraint signals satisfy $f_k\leq\bar{f}_k$ and $f_N\leq\bar{f}_N$ for all $w_k \in \mathcal{W}_k$ and all $\Delta_k\in\bm\B\pmb{\Delta}$ and for all $k\in\N$ . 
 The RMPC problem can be posed as a min-max problem~\cite{Scokaert98}, where the objective is to find a feasible $(\hat{K}_0,\hat{K}, \hat{\upsilon})$ that solves
\begin{equation}\label{eq:minmaxcostfun}
\mathbf{J}= \min_{(\hat{K}_0, \hat{K}, \hat{\upsilon}) \in \mathcal{U}} \hspace{0.2cm} \max_{\hat\Delta\in \bm\B\pmb{\Delta}} \mathcal{Z}(\hat{K}_0,\hat{K},\hat{\upsilon},\hat\Delta),
\end{equation}
where $\mathcal{U}$ is defined to be the set of all feasible control variables ($\hat{K}_0, \hat{K},\hat{\upsilon})$ such that all the problem constraints are satisfied:
$$
\mathcal{U}\!:=\! \{ ([\hat{K}_0~\hat{K}],\hat{\upsilon})\!\in\!\mathcal{K}\!\times\!\mathbb{R}^{N_u}\!\!:\!\mathcal{F}(\hat{K}_0,\hat{K},\hat{\upsilon},\hat\Delta)\!\leq \!\bar{\mathbf{f}},
       \forall \hat\Delta \!\in\! \bm\B\pmb{\hat\Delta}\}.
$$
$K_0,~K$ and $\pmb{\upsilon}$ can be computed online and applied in the usual receding horizon MPC manner, where the first input of the control sequence $\mathbf{u}$ is applied to the plant, the time window is shifted by 1, the current state is read and the process is repeated.
Since the optimization in (\ref{eq:minmaxcostfun}) is nonconvex, a semidefinite relaxation is used by introducing an upper bound ${\gamma^2}$ on the cost function. Using Lemma~\ref{Lemma1} and a Schur complement argument, the next result derives nonlinear conditions for solving (\ref{eq:minmaxcostfun}).
\begin{thm}\cite{Tahir13}\label{thm:NLMI}
Let all the variables be defined as above. Then $\mathcal{Z}(\hat{K}_0,\hat{K},\hat{\upsilon},\hat\Delta)\leq {\gamma^2} $ and $\mathcal{F}(\hat{K}_0,\hat{K},\hat{\upsilon},\hat\Delta)\le\bar{\mathbf{f}}$ for all $\hat\Delta \!\in\! \bm\B\pmb{\hat\Delta}$ if there exists a solution to the nonlinear matrix inequalities
\begin{eqnarray}\label{cost_LMI}
  T_1+\H(T_2\hat{K}\mathbf{B}_p T_3)&\succ& 0,\\
  \label{constr_LMI}
    T^i_1+\H(T^i_2\hat{K}\mathbf{B}_p T^i_3)&\succ& 0,~i=1,\ldots,N_f,
\end{eqnarray}
\end{thm}
where $\begin{bmatrix}T^i_1&\!\!\!\!\!T^i_2\\
   T^i_3&\!\!\!\!\!0\end{bmatrix}\!\!=\!\! \!$
   \begin{equation*}
   \!\!\!\!\!\!\!\!\!\!
   \left.\begin{array}{rl}~&\!\!\left.\!\!\!\!\!\!\!\!\!\begin{array}{ccccc}~~~~\scriptstyle{1}&~~~~~~~~~~~~~~~~~~~~\scriptstyle{N_q}&~~~~~~~~~~~~~\scriptstyle{N_p}&~~~~~~~\scriptstyle{N_u}\end{array}\right.\\
\left.\begin{array}{c}\scriptstyle{1}\\\scriptstyle{N_q}\\\scriptstyle{N_p}\\\scriptstyle{N_p}\end{array}\right.&\!\!\!\!\!\!\!\!\!\!\!\!\!\!
\left[\!\!\!\!\begin{array}{ccc|c}
    e_i^T(\bar{\mathbf{f}}\!-\!\mathbf{D}_{f}^{\hat{K}_0,\hat{\upsilon}}) &\!\!\!\!\! (\mathbf{D}_{q}^{\hat{K}_0,\hat{\upsilon}})^T\!\!\!-\!\frac{e_i^T}{2}\mathbf{D}_{fp} G_i^T&\!\!\!\! -\frac{e_i^T}{2}\mathbf{D}_{fp} S_i&\!\!\!\! -\frac{e_i^T}{2}\mathbf{D}_{fu}\\
    \ast &\!\!\!\!  R_i\!+\!\H\!\left(\mathbf{D}_{qp}G_i^T\right)&\!\!\!\! \mathbf{D}_{qp} S_i&\!\!\!\!\mathbf{D}_{qu}\\
    \ast &\!\!\!\! \ast &\!\!\!\! S_i&\!\!\!\!0\\\hline 0&\!\!\!\!G_i^T&\!\!\!\!S_i&\!\!\!\!0\end{array}\!\!\!\!\right]\!\!\!\end{array}\right.\!\!\!\!\!\!\!,
\end{equation*}
\begin{equation*}
\begin{bmatrix}T_1&\!\!T_2\\T_3&\!\!0\end{bmatrix}\!=\!\!\!\!\!\!\!\!\!\!
   \left.\begin{array}{rl}~&\!\!\left.\!\!\!\!\!\!\!\!\!\begin{array}{ccccc}~\!\scriptstyle{N_z}&~~\scriptstyle{1}&~~~~~~~~~~~\scriptstyle{N_q}&~~~~~~~~~\scriptstyle{N_p}&~~\!\scriptstyle{N_u}\end{array}\right.\\
\left.\begin{array}{c}\scriptstyle{N_z}\\\scriptstyle{1}\\\scriptstyle{N_q}\\\scriptstyle{N_p}\\\scriptstyle{N_p}\end{array}\right.&\!\!\!\!\!\!\!\!\!\!\!\!\!\!
\left[\begin{array}{cccc|c}
    I & \!\!\mathbf{D}_{z}^{\hat{K}_0, \hat{\upsilon}} &\!\! \mathbf{D}_{zp} G^T &\!\! \mathbf{D}_{zp} S&\!\!\mathbf{D}_{zu}\\
    \ast &\!\! {\gamma^2} &\!\! (\mathbf{D}_{q}^{\hat{K}_0,\hat{\upsilon}})^T &\!\! 0&\!\!0 \\
    \ast &\!\! \ast &\!\! R\!+\!\H\!\left(\mathbf{D}_{qp}G^T\right) &\!\! \mathbf{D}_{qp}S&\!\!\mathbf{D}_{qu}\\
    \ast &\!\! \ast &\!\! \ast &\!\! S&\!\!0\\\hline
    0&\!\!0&\!\!G^T&\!\!S&\!\!0
    \end{array}\right]\end{array}\right.\!\!\!\!\!\!\!\!,
    \end{equation*}
where $([ \hat{K}_0 ~ \hat{K}],\hat{\upsilon})\!\in\!\mathcal{K}\times\mathbb{R}^{N_u}$ and  $(S,R,G),\:(S_i,R_i,G_i)\!\in\!\hat{\Psi}$,  $i \!\in\! \mathcal{N}_f\!:=\!\{1,\ldots,N_f\}$ are slack variables with $\hat{\Psi}$ defined in \eqref{Slack_variable}.
In the sequel, we will occasionally write $T_1(\gamma^2,\hat{K}_0,\hat{\upsilon},S,R,G)$ etc. to emphasise dependence on the variables. It follows that the relaxed RMPC problem can be summarized as:
 \begin{equation} \label{finalprob}
 \begin{aligned}
\hspace{-2mm}\min\{  {\gamma^2} \!:&([ \hat{K}_0 ~~ \hat{K}],\hat{\upsilon})\!\in\!\mathcal{K}\!\times\mathbb{R}^{N_u},(\ref{cost_LMI}),(\ref{constr_LMI}){\rm~are~satisfied},\\
&(S,R,G),( S_i, R_i, G_i)\!\in\! \widehat{\Psi},\: i \!\in\! \mathcal{N}_f\}.
\end{aligned}
\end{equation}
Definitions (\ref{def1})-(\ref{def2}) verify that (\ref{finalprob}) is nonlinear due to terms of the form $\K \mathbf{B}_p \Phi^T$ where $\Phi$ stands for $S,S_i,G$ and $G_i$. Note that (\ref{finalprob}) is linear for fixed $K$ and RMPC schemes with fixed $K$ have been proposed \cite{Chisci02}. However, this introduces conservatism depending on the choice of $K$. A linearization scheme is proposed in \cite{Tahir13}, which uses an S-procedure to separate $\hat{K}$. However, this scheme has a high computational burden. Furthermore, some of the introduced linearization variables are restricted to a specific form. To overcome these two limitations, a new linearization procedure for (\ref{finalprob}) is proposed, which substantially reduces the computational complexity at the expense of only minor conservatism in the formulation.
\section{Linearization scheme for the relaxed RMPC problem}\label{sec:Linearization}
LMI-based methods reduce conservatism by explicitly incorporating uncertainty within the online optimization. However, they suffer from a heavy computational cost, which makes them impractical for systems with a high number of states or fast dynamics. In this section, a novel linearization procedure  is proposed to overcome the nonconvexity for the RMPC problem in (\ref{finalprob}), while conservativeness and computation burden are kept at low levels.
We will use the following form of the Elimination Lemma.
\begin{lem}[Elimination Lemma]\label{Elimination}
Let $Q\!=\!Q^T\!\in\! \mathbb{R}^{n\times n},B\! \in \!\mathbb{R}^{n\times m}$ and $C\! \in\! \mathbb{R}^{n\times p}$ be given matrices and let $B^{\perp}$ and $C^{\perp}$ denote orthogonal complements of $B$ and $C$, respectively. Then the following two statements are equivalent: (i) $(B^{\perp})^T\! Q (B^{\perp}) \succ 0 ~\&~(C^{\perp})^T\! Q (C^{\perp}) \succ 0$. (ii) $\exists ~ Z \in \mathbb{R}^{p\times m}: ~~ Q+\H(C Z B^T) \succ 0$.
\end{lem}
The proof and some applications of the Elimination Lemma can be found in \cite{Skelton97,Hu20}. The next result uses the Elimination Lemma to derive LMI sufficient conditions for the nonlinear matrix inequality conditions of Theorem~\ref{thm:NLMI}.
\begin{thm} \label{New_linearization}
Let all variables be as defined Section~\ref{sec:Problem}. Then,
$\mathcal{Z}(\K_0,\K,\hat{\upsilon},\hat{\Delta})\le{\gamma}^2$ and $\mathcal{F}(\hat{K}_0,\hat{K},\hat{\upsilon},\hat\Delta)\le\bar{\mathbf{f}}$ for all 
$\hat{\Delta}\!\in\!\bm\B\pmb{\hat{\Delta}}$  if there exist
solutions $([\K_0~~\K],\hat{\upsilon})\in\mathcal{K}\!\times\mathbb{R}^{N_u}$, $X \in \mathbb{R}^{N_n \times N_n}$, with $X$ lower block-diagonal with $n\times n$ blocks, $(S,R,G),~( S_i,R_i,G_i) \in \widehat{ \Psi},~\forall i \in \mathcal
N_f$ to the following LMIs:
\begin{eqnarray}\label{linear_cost_LMI}
    \begin{bmatrix}
    T_1+\H\left(T_2\bar{K}Y^*\right) & \ast\\
    \left(\mathbf{B}_p T_3-\bar{K}^TT_2^T\right)-XY^* & X+X^T
    \end{bmatrix}\!\!&\succ&\! 0\\\label{linear_constr_LMI}
    \begin{bmatrix}
    T^i_1+\H\left(T^i_2\bar{K}Y^*_i\right) & \ast\\
    \left(\mathbf{B}_p T^i_3-\bar{K}^T(T^i_2)^T\right)-XY^*_i & X+X^T
    \end{bmatrix}\!\!&\succ&\! 0,
\end{eqnarray}
for some $Y^*\in\R^{N_n\times(N_z+1+N_q+N_p)},Y^*_i\in\R^{N_n\times(1+N_q+N_p)}$ and where $\bar{K} := \K X\in\mathcal{K}$. Furthermore, suppose that \eqref{cost_LMI} and \eqref{constr_LMI} have  feasible solutions for $(\gamma^2,\K_0,\K,\hat{\upsilon},R,S,G,R_i,S_i,G_i)=(\gamma^{2*},\K_0^*,\K^*,\hat{\upsilon}^*,R^*,S^*,G^*,R_i^*,S_i^*,G_i^*)$ so that 
\begin{eqnarray}
\label{T10}   \hspace{-4mm}T_1(\gamma^{2*},\K_0^*,\hat{\upsilon}^*,S^*,R^*,G^*)\!+\!\H\left(T_2\hat{K}^*\mathbf{B}_p T_3(S^*,G^*)\right)\!\!\!\!&\!\!\!\!\succ\!\!\!\!&\!\!\!\!0,\\\nonumber
    T^i_1(\K_0^*,\hat{\upsilon}^*,R_i^*,S_i^*,G_i^*)+\H\left(T^i_2\hat{K}^*\mathbf{B}_p T^i_3(S_i^*,G_i^*)\right)\!\!\!\!&\!\!\!\!\succ\!\!\!\!&\!\!\!\! 0,
\end{eqnarray}
and let $Y^*=\mathbf{B}_pT_3(S^*,G^*)+(T_2\K^*)^T$ and $Y^*_i=\mathbf{B}_pT^i_3(S^*_i,G^*_i)+(T^i_2\K^*)^T$. Then \eqref{linear_cost_LMI} and \eqref{linear_constr_LMI} are feasible. 
\end{thm}
\begin{proof}
We prove the first part by proving that the LMIs in \eqref{linear_cost_LMI} and \eqref{linear_constr_LMI} are sufficient for the nonlinear matrix inequalities in \eqref{cost_LMI} and \eqref{constr_LMI}, respectively. We first use the Elimination Lemma to give an equivalent form to \eqref{cost_LMI}. 
In order to separate $\hat{K}$ from $T_3$, the inequality in (\ref{cost_LMI}) can be rearranged as:\vspace{-2mm}
\begin{equation}\label{separation}
    \begin{bmatrix}
    I & T_2 \hat{K}
    \end{bmatrix} \overbrace{\begin{bmatrix}
    T_1 & T_3^T\mathbf{B}_p^T \\
    \mathbf{B}_p T_3 & 0
    \end{bmatrix}}^{Q} \overbrace{\begin{bmatrix}
    I \\ \hat{K}^T T_2^T
    \end{bmatrix}}^{C^\perp}\succ 0.
\end{equation}
Then, applying the Elimination Lemma~\ref{Elimination} on (\ref{separation}) (with $B=I$) shows that \eqref{separation}, hence \eqref{cost_LMI} is equivalent to
\begin{equation}\label{Elm_Ineq}
    \begin{bmatrix}
    T_1 &  \!\!\!\!T_3^T\mathbf{B}_p^T\\
    \mathbf{B}_p T_3 & \!\!\!\!0
    \end{bmatrix}\!\! +\!\! \overbrace{\begin{bmatrix}
     -T_2 \hat{K} \\ I
    \end{bmatrix}}^{C} \overbrace{\begin{bmatrix}Y&\!\!\!\!X\end{bmatrix}}^{Z} \!\!+\!\! \begin{bmatrix}Y^T\\X^T\end{bmatrix} \begin{bmatrix}
    -\hat{K}^T T_2^T & \!\!\!\!I
    \end{bmatrix}\!\!\succ\! 0,
\end{equation}
where $Y$ and $X$ are free slack variables. Since $\H(X)\!\succ\!0$, $X$ is nonsingular and we can define $\bar{K}\!:=\!\hat{K}X$ as a new variable. To preserve the structure of $\hat{K}$ which ensures causality, we restrict $X$ to be block lower triangular (with $n\times n$ blocks). To preserve linearity, we restrict $Y$ to have the form $Y\!=\!-XY^*$ with $Y^*$ free (but not a variable). Substituting $Y\!=\!-XY^*$ into \eqref{Elm_Ineq} proves that \eqref{linear_cost_LMI} is sufficient for \eqref{cost_LMI} (but not necessary due to the restrictions on $Y$ and $X$). A similar procedure proves that \eqref{linear_constr_LMI} are sufficient for \eqref{constr_LMI}.
Next, we prove feasibility of \eqref{linear_cost_LMI} and \eqref{linear_constr_LMI}. To show that \eqref{linear_cost_LMI} has a feasible solution, set $(\gamma^2,\K_0,\K, \hat{\upsilon},R,S,G,X)\!=\!(\gamma^{2*}, \K^*_0,\K^*,\hat{\upsilon}^*,R^*,S^*,G^*,I)$. Then the LHS of \eqref{linear_cost_LMI} becomes
\bean
    T^*:=\begin{bmatrix}
    T_1^*+\H\left(T_2\K^*\left(\mathbf{B}_pT_3^*+(\K^*)^TT_2^T\right)\right) & \ast\\
    -2(\K^*)^TT_2^T & 2I
    \end{bmatrix}
\eean
where $T_1^*\!:=\!T_1(\gamma^{2*},\K_0^*,\hat{\upsilon}^*,S^*,R^*,G^*)$ and $T_3^*\!:=\!T_3(S^*,G^*)$. Applying a Schur complement on $T^*$ shows that $T^*\!\succ\!0$ if and only if \eqref{T10} is satisfied. It follows that \eqref{linear_cost_LMI} is feasible if \eqref{T10} is. 
A similar procedure proves the feasibility of \eqref{linear_constr_LMI}.
\end{proof}
\begin{rem}
Theorem~\ref{New_linearization} provides sufficient LMI conditions for the initial nonconvex RMPC problem. Therefore, $K_0,~K$ and $\upsilon$ can be computed online and applied in a receding horizon manner as stated in Section~\ref{RMPC_problem}.
\end{rem}
\begin{rem}\label{rem:N_linearization}
In comparison to \cite{Tahir13}, the novelty of the proposed linearization procedure is that it does not restrict the structure of the slack variables $(R,S,G)$ and $(R_i,S_i,G_i)$ beyond the requirements of $\widehat{ \Psi}$, and therefore it is less conservative.
\end{rem}
\section{Single LMI approach for handling constraints signal for RMPC problem}\label{sec:SingleLMI}
Instead of solving multiple matrix inequalities for the constraints (one for each of the $N_f$ constraints \eqref{constr_LMI} or \eqref{linear_constr_LMI}), we propose a strategy to combine all within a single inequality. This results in reduced computational complexity and improved algorithm scalability. Our algorithm is based on the following result which uses an S-procedure to derive one LMI condition that is sufficient for a set of elementwise inequalities. 

\begin{thm}\label{Elementwise}
Let $\tilde{f}\in\R^{N_f}$ and let $e\in\R^{N_f}$ be the vector of ones. Then $\tilde{f}\geq 0$ if there exist $\mu\in\R$ and ${M}\in\D^{N_f}$ such that,
\begin{equation}\label{L}
    \mathcal{L}:=\left[\begin{array}{cc}2\mu&\left(\tilde{f}-{M}e-e\mu \right)^T\\\ast&{M}+{M}^T\end{array}\right]\succeq0.
\end{equation}
\end{thm}
\vspace{2mm}
\begin{proof}
Let $\pmb{\Omega}\!:=\!\{\diag(\d_1,\ldots,\d_{N_f})\!:\d_i\!\in\!\{0,1\},\sum\limits_{i=1}^{N_f}\d_i\!=\!1\}$. Then, 
\begin{equation}\label{ft}
    \tilde{f}\ge0\Leftrightarrow e^T\Delta\tilde{f}+\tilde{f}^T\Delta^Te\ge0~\forall\Delta\in\pmb{\Omega}.
\end{equation}
Let $\Delta\!\in\!\pmb{\Omega}$. Since $\d_i\!\in\!\{0,1\}$ and $\sum_{i\!=\!1}^{N_f}\d_i\!=\!1$, then
\begin{equation}\label{Smu}
\begin{split}
& M_{\Delta}:=\Delta {M}\!+\!{M}^T\Delta^T\!-\!\Delta({M}\!+\!{M}^T)\Delta^T\!=\!0~\forall {M}\!\in\!\D^{N_f},\\
& \mu_{\Delta}:=e^T\Delta e\mu\!+\!\mu e^T\Delta^Te\!-\!2\mu\!=\!0~\forall \mu\!\in\!\R,
\end{split}
\end{equation}
respectively. It is straightforward to verify the identity,
\begin{equation*}
e^T\Delta\tilde{f}\!+\!\tilde{f}^T\Delta^Te \!=\! e^TM_{\Delta}e\!+\!\mu_{\Delta}\!+\!\left[\begin{array}{cc}1&e^T\Delta\end{array}\right]\mathcal{L}\left[\begin{array}{c}1\\\Delta^Te\end{array}\right]\!\!\cdot
\end{equation*}
The proof now follows from \eqref{ft} and \eqref{Smu}.
\end{proof}

Theorem~\ref{Elementwise} enables us to give sufficient conditions for the constraints in \eqref{constr_LMI} in the form of a single matrix inequality.

\begin{thm}\label{SngLmiThm}
Let all variables be as defined Section~\ref{sec:Problem}. Then,
$\mathcal{Z}(\K_0,\K,\hat{\upsilon},\hat{\Delta})\le{\gamma}^2$ and $\mathcal{F}(\hat{K}_0,\hat{K},\hat{\upsilon},\hat\Delta)\le\bar{\mathbf{f}}$ for all 
$\hat{\Delta}\!\in\!\bm\B\pmb{\hat{\Delta}}$  if there exist
solutions $([\K_0~~\K],\hat{\upsilon})\in\mathcal{K}\!\times\mathbb{R}^{N_u}$, $(S,R,G), ( \tilde{S}, \tilde{R}, \tilde{G}) \in \widehat{ \Psi}$, $\mu\in\R$ and $M\in\D^{N_f}$
to \eqref{cost_LMI} and,
\begin{equation}\label{stateupperi0}
    \tilde{T}_1+\H(\tilde{T}_2\hat{K}\mathbf{B}_p \tilde{T}_3)\succ 0,
\end{equation}
where $\begin{bmatrix}\tilde{T}_1&\!\!\tilde{T}_2\\\tilde{T}_3&\!\!0\end{bmatrix}\!=$\vspace{-2mm}
\begin{equation*}
    \left.\begin{array}{rl}~&\!\!\left.\!\!\!\!\!\!\!\!\!\!\!\!\!\!\!\!\!\!\begin{array}{ccccc}~~~\scriptstyle{1}&~~~~~~~~\scriptstyle{N_f}&~~~~~~~~~~~~~~~\scriptstyle{N_q}&~~~~~~~~~~~\scriptstyle{N_p}&~~~~\scriptstyle{N_u}\end{array}\right.\\
\!\!\!\!\!\!\left.\begin{array}{c}\scriptstyle{1}\\\scriptstyle{N_f}\\\scriptstyle{N_q}\\\scriptstyle{N_p}\\\scriptstyle{N_p}\end{array}\right.&\!\!\!\!\!\!\!\!\!\!\!\!\!\!
\left[\!\!\!\!\!\begin{array}{cccc|c}
    2\mu & \!\!\!\!\!\!\!(\bar{\mathbf{f}}\!\!-\!\mathbf{D}_f^{\K_0,\hat{\upsilon}}\!\!-\!\!Me\!-\!e\mu)^T &\!\!\!\!\! \!\!\!\!\!\!\!(\mathbf{D}_q^{\K_0,\hat{\upsilon}})^T &\!\!\!\!\!\!\! 0&\!\!\!\!\!0\\
    \ast &\!\!\!\!\!\!\! M\!+\!M^T &\!\!\!\!\!\!\!\!\! -\mathbf{D}_{fp}\tilde{G}^T &\!\!\!\!\!\!\! -\mathbf{D}_{fp}\tilde{S}&\!\!\!\!\!-\mathbf{D}_{fu} \\
    \ast &\!\!\!\!\!\!\! \ast &\!\!\!\!\!\!\!\!\!\!\!\! \tilde{R}\!+\!\H\!(\mathbf{D}_{qp}\tilde{G}^T) &\!\!\!\!\!\!\! \mathbf{D}_{qp}\tilde{S}&\!\!\!\!\!\mathbf{D}_{qu}\\
    \ast &\!\!\!\!\!\!\! \ast &\!\!\!\!\!\!\!\!\!\!\!\! \ast &\!\!\!\!\!\!\! \tilde{S}&\!\!\!\!\!0\\\hline
    0&\!\!\!\!\!\!\!0&\!\!\!\!\!\!\!\!\!\!\!\!\tilde{G}^T&\!\!\!\!\!\!\!\tilde{S}&\!\!\!\!\!0
    \end{array}\!\!\!\!\!\right]\end{array}\right.\!\!\!\!\!\!\!\!\!\cdot
    \end{equation*}
\end{thm}

\begin{proof}
 We only need to prove that \eqref{stateupperi0} is sufficient for $\tilde{f}\!:=\!\bar{\mathbf{f}} \!-\! \mathbf{f}
\! \! \geq\! 0$ for all $\hat{\Delta}\!\in\!\bm\B\pmb{\hat{\Delta}}$, where $\mathbf{f}$ is defined in (\ref{def2}). Using Theorem~\ref{Elementwise} and rearranging \eqref{L} verifies that a sufficient condition for the constraints is
\begin{equation} \label{xbar10}
H_{11}\!+\!\H(H_{12}\hat\Delta(I\!-\!H_{22}\hat\Delta)^{-1}\!H_{21})\! \!\succ \! 0,~\forall\hat{\Delta}\!\in\!\bm\B\pmb{\hat{\Delta}},
\end{equation}
where we have used a strict inequality to avoid issues related to ill-conditioning near optimality and where 

\bean
\left[\!\!\begin{array}{c|c}H_{11}&H_{12}\\\hline
H_{21}&H_{22}\end{array}\!\!\right]\!\!:=\!\!\left[\!\!\begin{array}{cc|c}2\mu&\left(\bar{\mathbf{f}}\!-\!\mathbf{D}_{f}^{\K_0,\hat{\upsilon}}\!-\!Me\!-\!\mu e\right)^T &0\\
\ast&M\!+\!M^T&-\mathbf{D}_{fp}^\K \\ \hline \mathbf{D}_{\hat{q}}^{\K_0,\hat{\upsilon}}&0& \mathbf{D}_{qp}^\K \end{array}\!\!\right]\!\!\!\cdot \eean Using Lemma~\ref{Lemma1}
on (\ref{xbar10}) and the definition of (\ref{def1}) yields the matrix inequality (\ref{stateupperi0}) as a sufficient condition.
\end{proof}
Using the linearization procedure in Section~\ref{sec:Linearization}, we next derive sufficient LMI conditions for the problem stated in (\ref{finalprob}).

\vspace{2mm}
\begin{thm}\label{Single_LMI}
Let all variables be as defined Theorem~\ref{SngLmiThm}. Then,
$\mathcal{Z}(\K_0,\K,\hat{\upsilon},\hat{\Delta})\!\le\!{\gamma}^2$ and $\mathcal{F}(\hat{K}_0,\hat{K},\hat{\upsilon},\hat\Delta)\!\le\!\bar{\mathbf{f}}$ for all 
$\hat{\Delta}\!\in\!\bm\B\pmb{\hat{\Delta}}$  if there exist $([\K_0~\K],\hat{\upsilon})\!\in\!\mathcal{K}\!\times\mathbb{R}^{N_u}$, $(S,R,G), ( \tilde{S}, \tilde{R}, \tilde{G}) \!\in\! \widehat{ \Psi}$, $\mu\!\in\!\R$, $M\!\in\!\D^{N_f}$ and $X \!\in\! \mathbb{R}^{N_n \times N_n}$, with $X$ lower block-triangular with $n\times n$ blocks, to (\ref{linear_cost_LMI}) and the following LMI: 
\begin{equation}\label{linear_constr_Single_LMI}
    \begin{bmatrix}
    \tilde{T}_1+\H\left(\tilde{T}_2\bar{K}\tilde{Y}^*\right) & \ast\\
    \left(\mathbf{B}_p \tilde{T}_3-\bar{K}^T\tilde{T}_2^T\right)-X\tilde{Y}^* & X+X^T
    \end{bmatrix}\!\!\succ\! 0,
\end{equation}
for any $Y^*\!\in\!\R^{N_n\times(N_z\!+\!1\!+\!N_q\!+\!N_p)},~\tilde{Y}^*\!\in\!\R^{N_n\times(1\!+\!N_f\!+\!N_q\!+\!N_p)}$ and where $\bar{K} := \K X\in\mathcal{K}$. Furthermore, suppose that \eqref{cost_LMI} and \eqref{stateupperi0} have  feasible solutions for $(\gamma^2,\K_0,\K,\hat{\upsilon},R,S,G,\tilde{R},\tilde{S},\tilde{G})=(\gamma^{2*},\K^*_0,\K^*,\hat{\upsilon}^*,R^*,S^*,G^*,\tilde{R}^*,\tilde{S}^*,\tilde{G}^*)$ so that \eqref{T10} and 
\begin{equation}\label{T1t0}
    \tilde{T}_1(\K^*_0,\K^*,\hat{\upsilon}^*,\tilde{R}^*,\tilde{S}^*,\tilde{G}^*)+\H\left(\tilde{T}_2\hat{K}^*\mathbf{B}_p \tilde{T}_3(\tilde{S}^*,\tilde{G}^*)\right) \succ 0.
\end{equation}
are satisfied and let $Y^*=\mathbf{B}_pT_3(S^*,G^*)+(T_2\K^*)^T$ and $\tilde{Y}^*=\mathbf{B}_p\tilde{T}_3(\tilde{S}^*,\tilde{G}^*)+(\tilde{T}_2\K^*)^T$. Then \eqref{linear_cost_LMI} and \eqref{linear_constr_Single_LMI} are feasible.

\end{thm}
\vspace{2mm}
\begin{proof} The result can be proved by applying the Elimination Lemma~\ref{Elimination} on  \eqref{stateupperi0} in a similar procedure to that used in the proof of Theorem~\ref{New_linearization} and is omitted.
\end{proof}

\begin{rem}\label{rem:Additional _conservativeness}
Note that \eqref{L} gives only sufficient conditions for $\tilde{f}\!\ge\!0$ and can be conservative. To reduce the conservativeness, we can add redundant constraints, e.g.  $\sum_{i=1}^{N_f}\d_i^2\!=\!1$. 
However, this is not pursued here. Our numerical experimentation, including the examples below, indicates that the single LMI sufficient condition provided by Theorem~\ref{Single_LMI} performs as well as the multiple LMI conditions provided by Theorem~\ref{New_linearization}. 
\end{rem}

\section{Feasibility analysis}\label{sec:Feasibility}

A major problem in MPC is to ensure that the constraints are feasible. Infeasibility may arise if the constraints are too tight or it may be due to the approximations used to obtain a practical solution. In the context of this work, to guarantee feasibility, Theorems~\ref{New_linearization} and \ref{Single_LMI} require initial feasible solutions to \eqref{cost_LMI}, and \eqref{constr_LMI} (to compute $Y^*$ and $Y^*_i$) or \eqref{cost_LMI} and \eqref{stateupperi0} (to compute $Y^*$ and $\tilde{Y}^*$). On the other hand, \eqref{cost_LMI} and \eqref{constr_LMI} are nonlinear and difficult to solve and these computations need to be carried out online. In this section we develop algorithms that address these issues that involve extensive computations, which, however, are convex and can be carried out offline.  We will concentrate on Theorem~\ref{Single_LMI} since the procedure for Theorem~\ref{New_linearization} is similar.

Our approach is to find solutions to \eqref{cost_LMI} and \eqref{stateupperi0} offline that are feasible for every $x_0$ in a constrained set. Note that both \eqref{cost_LMI} and \eqref{stateupperi0} can be written as $ M(x_0) \!:=\! M_1 \!+\!\H (M_2 x_0 M_3)\! \succ 0\!$ in which $M_3$ is constant and $M_1$ and $M_2$ are independent of $x_0$, see \eqref{def1}. The next result uses an S-Procedure to derive sufficient conditions for $ M(x_0)\! \succ\!0$ for all $x_0$ in a polytopic set and forms the basis for our Algorithm~\ref{alg_1}.

\begin{thm}\label{All_x0}
Let $M_1=M_1^T\in\R^{m\times m},~M_2\in\R^{m\times n},~M_3\in\R^{1\times m},~ C_0\in\R^{p\times n},~\underline{c}_0\le\bar{c}_0\in\R^p$ be given. Then $M_1+\H(M_2x_0M_3)\succ0$ for all $x_0\in\mathcal{X}_0:=\{x_0\in\R^n:\underline{c}_0\le C_0x_0\le\bar{c}_0\}$ if there exists $0\preceq D_0\in\D^p$ such that
\bean
L:=\begin{bmatrix}M_1+\frac{1}{2}\H(M_3^T(\underline{c}_0^TD_0\bar{c}_0)M_3)&\ast\\
    M_2^T-\frac{1}{2}C_0^TD_0(\underline{c}_0+\bar{c}_0)M_3&C_0^TD_0C_0\end{bmatrix}\succ0.
\eean
\end{thm}
\vspace{2mm}
\begin{proof}
A manipulation verifies the following identity
\begin{equation*}
    M_1+\H(M_2x_0M_3)=M_0
    +\begin{bmatrix}I_m&M_3^Tx_0^T\end{bmatrix}L\begin{bmatrix}I_m\\x_0M_3\end{bmatrix}\!\!,
\end{equation*}
where $M_0\!:=\!M_3^T(C_0x_0\!-\!\underline{c}_0)^T\!D_0(\bar{c}_0\!-\!C_0x_0)M_3$. The result then follows from the constraints on $x_0$ and the structure and sign-definiteness of $D_0$ (which ensure that $M_0\!\succ\!0$ for all $x_0\in\mathcal{X}_0$) and since $L\succ0$ (which ensures that the second term on the RHS of the identity is positive definite for all $x_0\in\mathbb{R}^n$).
\end{proof}

Theorem \ref{All_x0} gives an LMI procedure for solving \eqref{cost_LMI} and \eqref{stateupperi0} for all $x_0\!\in\!\mathcal{X}_0$ when $\hat{K}$ is given and for solving
\eqref{linear_cost_LMI} and \eqref{linear_constr_Single_LMI} for all $x_0\!\in\!\mathcal{X}_0$ when an initial feasible solution for \eqref{cost_LMI} and \eqref{stateupperi0} is given. Algorithm~\ref{alg_1} outlines the suggested offline policy for computing initial feasible solutions for Theorem~\ref{Single_LMI}. 
\begin{algorithm}[ht!]\label{alg_1}
\SetAlgoLined
\KwResult{$Y^*(S^*,G^*,\K^*)$ and $\tilde{Y}^*(\tilde{S}^*,\tilde{G}^*,\K^*)$ }
 \textbf{Step 1:} \\
 In Theorem~\ref{SngLmiThm}, fix $\hat{K}$ (e.g. $\hat{K}\!=\!0$), replace $\bar{f}$ by $\beta\bar{f}$ and use Theorem~\ref{All_x0} to minimize $\beta$ such that \eqref{stateupperi0} is satisfied for all $x_0\!\in\!\mathcal{X}_0$. Record $\beta$, $\tilde{S}$ and $\tilde{G}$ and let $\K^*\!=\!0,\tilde{S}^*\!=\!\tilde{S},\tilde{G}^*\!=\!\tilde{G},i\!=\!1,\beta_i\!=\!\beta$. Select a maximum number of iterations $i_\max$ and tolerance $tol_{\beta}\!<\!1$\;
  \textbf{Step 2:} \\
  \While{($\beta>1$) \& ($i<i_\max$) }{ In Theorem~\ref{Single_LMI}, replace $\bar{f}$ by $\beta\bar{f}$ and use Theorem~\ref{All_x0} to find the smallest $\beta\ge1$ such that \eqref{linear_constr_Single_LMI} is satisfied for all $x_0\in\mathcal{X}_0$. Set $\beta_{i+1}=\beta$ and update $\hat{K}^*:=\hat{K}$, $\tilde{S}^*:=\tilde{S}$ and $\tilde{G}^*:=\tilde{G}$\;
  \If{($\frac{|\beta_{i+1}-\beta_i|}{\beta_{i+1}}<tol_{\beta}$)}{\textbf{break;} ~~~~(convergence to a $\beta>1$)}
  Set $i:=i+1$.}
  
 \textbf{Step 3:}\\
   \eIf{$\beta>1$}
   {Sub-divide $\mathcal{X}_0$ into smaller sets\;
   \textbf{Go back to Step 2}\;}
   {In Theorem~\ref{SngLmiThm} fix $\K\!=\!\K^*$ and use Theorem~\ref{All_x0} to minimize $\gamma^2$ such that \eqref{cost_LMI} is satisfied for all $x_0\in\mathcal{X}_0$. Record $\gamma^2$ and let $S^*=S$ and $G^*=G$\;}

\textbf{Step 4:} \\
  Set $j=1$, $\gamma^2_j=\gamma^2$ and select $j_\max$ to be the maximum number of iterations and $tol_{\gamma}<1$ to be a tolerance.\\
  \While{( ($j<j_\max$)}{
  In Theorem~\ref{Single_LMI}, use Theorem~\ref{All_x0} to minimize $\gamma^2$ such that \eqref{linear_cost_LMI} and~\eqref{linear_constr_Single_LMI} are satisfied for all $x_0\in\mathcal{X}_0$.
  Set $\gamma^2_{j+1}=\gamma^2$ and update $\K^*:=\K$, $S^*=S$, $G^*=G$, $\tilde{S}^*:=\tilde{S}$ and $\tilde{G}^*:=\tilde{G}$\;
    \If{($\frac{|\gamma^2_{j+1}-\gamma^2_j|}{\gamma^2_{j+1}}<tol_{\gamma}$)}{\textbf{break;}}
    Set $j:=j+1$.
 }
 \caption{Initial feasible solutions for Theorem~\ref{Single_LMI} }
\end{algorithm}

\begin{rem}
If $\beta\!=\!1$ at the end of Step 2, we have feasible solutions to \eqref{cost_LMI} and \eqref{stateupperi0} for all $x_0\!\in\!\mathcal{X}_0$ and we can use Theorem~\ref{Single_LMI} online. If fewer online computations are needed,  Theorem~\ref{SngLmiThm} can be used online with $\hat{K}\!=\!\hat{K}^*$ and $\K_0,\hat{\upsilon}$ can be used to minimize $\gamma$.
If $\beta\!>\!1$ at the end of Step 2, then, with minor modifications to Algorithm~\ref{alg_1}, Theorem~\ref{Single_LMI} can still be used, although without a guaranteed feasible solution, but possibly a good initial solution if $\beta\!\sim\!1$ since the conditions are not necessary. Alternatively, we may sub-divide $\mathcal{X}_0$ into subsets, find a feasible solution for each and use a look-up table to choose the initial solution given $x_0$ in the online implementation. A systematic procedure for sub-dividing $\mathcal{X}_0$ is under investigation, however, see Example~1 for more details.
\end{rem}
\begin{rem}\label{rec_feas}
Although this work concentrates on solving effectively the robust MPC control problem at the first iteration, some comments regarding recursive feasibility are given here. By selecting the terminal costs and constraints to lay inside a robust control invariant set, using a dual-mode controller similarly to \cite{Tahir13}, recursive feasibility and stability of the overall control scheme can be achieved. An alternative approach for recursive feasibility of the robust optimal control problem can be potentially proven by an adaptive horizon method similar to the work in \cite{Chen20}. However, further investigation into recursive feasibility and stability have been left for future work. 
\end{rem}
\section{Numerical Examples and Simulations}\label{sec:Alg_Example}
The effectiveness of our algorithms is illustrated by two examples. We use the CVX package~\cite{cvx}, in MATLAB R2019b on a computer with 2.40 GHz Intel Xeon(R) CPU and 64.0 GB memory.

\subsection{Example 1}
The system is a variation on that proposed in \cite{Tahir11,Tahir12,Tahir13}. It is a second order unstable process subject to uncertainty and bounded disturbances. The dynamics have the form of~(\ref{eq:dt_system_model}) with
\begin{equation*}
    A\!=\!C_q\!=\!\!\begin{bmatrix}
    1 & \!\!\!\!0.8 \\
    0.5 & \!\!\!\!1
    \end{bmatrix}\!\!,B_u\!=\!D_{qu}\!=\!\!\begin{bmatrix}
    1\\1
    \end{bmatrix}\!\!,B_w\!=\!\!\begin{bmatrix}
    0.1 \\ 0.1
    \end{bmatrix}\!\!,
    B_p\!=\!0.1I_2,\hat{C}_q\!=\!0.
\end{equation*}
The time-invariant uncertainty has the form $\pmb{\Delta}\!:=\!\{\delta I_2\!:\delta \!\in\! \mathbb{R}\}$ and the disturbances set is $\mathcal{W} \!:=\! \{ w \!\in\! \mathbb{R}^{n_w} : -1\!\leq\! w\leq\! 1\}$.
The constraints are $-\bar{u}\!\le\! u_k\!\le\!\bar{u}\!=\!8,k\!=\!0,\ldots,N\!-\!1$ and $-\bar{x}\!\le\! x_k\!\le\!\bar{x}\!=[7~7]^T,k\!=\!0,\ldots,N$, respectively and the initial state is set at the state constraints  boundary $x_0\!=\!\bar{x}$. The control objective is to regulate the unstable system subject to uncertainties and disturbances into the origin whilst satisfying the input and state constraints. The states and inputs are equally weighted in the cost function. Two RMPC algorithms are used. The first, {CE-RMPC\textit{\#1}$(\K_0,\K,\hat{\upsilon})$}, is described by Theorem~\ref{Single_LMI}, where the initial feasible solutions $(Y^*,\tilde{Y}^*)$ are computed offline by Algorithm~\ref{alg_1}. The second algorithm, CE-RMPC\textit{\#2}$(\K_0,\hat{\upsilon})$, is described by Theorem~\ref{SngLmiThm}, where $\K\!=\!\K^*$ (computed offline by Algorithm~\ref{alg_1}). For this example, $\beta$ is greater than 1 if we take $\mathcal{X}_0$ to be the entire constrained state-space ($\mathcal{X}_0\!:=\!\{x_0\!:-\bar{x}\!\le \!x_0\!\le\!\bar{x}\}$). Thus  $\mathcal{X}_0$ is divided into 25 smaller sets $\mathcal{X}_0^{i,j}$, with $\beta\!=\!1$ for each of these sets, and a look-up table has been used to store $(Y^*,\tilde{Y}^*)$ for each subset. The subsets used are
 \bean
 \mathcal{X}_0^{i,j}\!=\!\{\!\left[\!\!\begin{array}{c}x_{01}\\x_{02}\end{array}\!\!\right]\!\!\!:\underline{x}_{01}^i\!\le \!x_{01}\!\le\!\bar{x}_{01}^i,~\underline{x}_{02}^j\!\le\! x_{02}\!\le\!\bar{x}_{02}^j\},i,j\!=\!1,\ldots,5,
 \eean
 where for $k=1,2,$
 \bean\begin{aligned}
 &[\underline{x}_{0k}^1,\bar{x}_{0k}^1]\!=\!7[.75,1],~[\underline{x}_{0k}^2,\bar{x}_{0k}^2]\!=\!7[.4,.75],~ [\underline{x}_{0k}^3,\bar{x}_{0k}^3]\!=\!7[-.4,.4],\\
 &[\underline{x}_{0k}^4,\bar{x}_{0k}^4]\!=\!-7[.75,.4],~
 [\underline{x}_{0k}^5,\bar{x}_{0k}^5]\!=\!-7[1,.75].
 \end{aligned}
 \eean
The time to compute the initial solutions $(Y^*,\tilde{Y}^*)$ for all subsets is 94.5 seconds. Note that offline computation time depends on the number of subsets and terminal iteration values $i_{max}$ and $j_{max}$. The controller from \cite{Tahir13} is also presented for comparison. All algorithms are simulated with prediction horizon $N\!=\!5$.
 
\begin{figure}[h]
\centering
\includegraphics[width=\columnwidth]{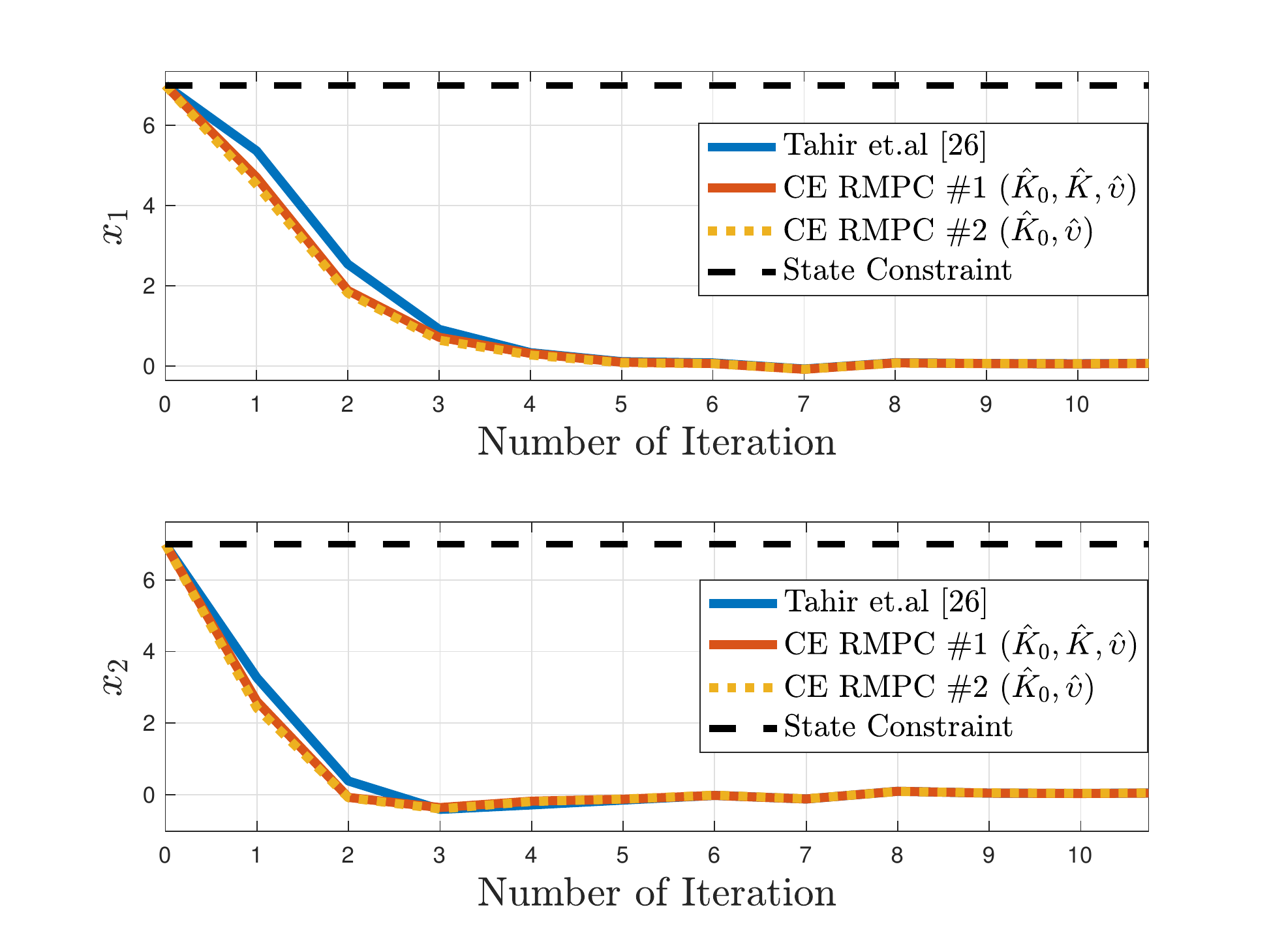}
\includegraphics[width=\columnwidth]{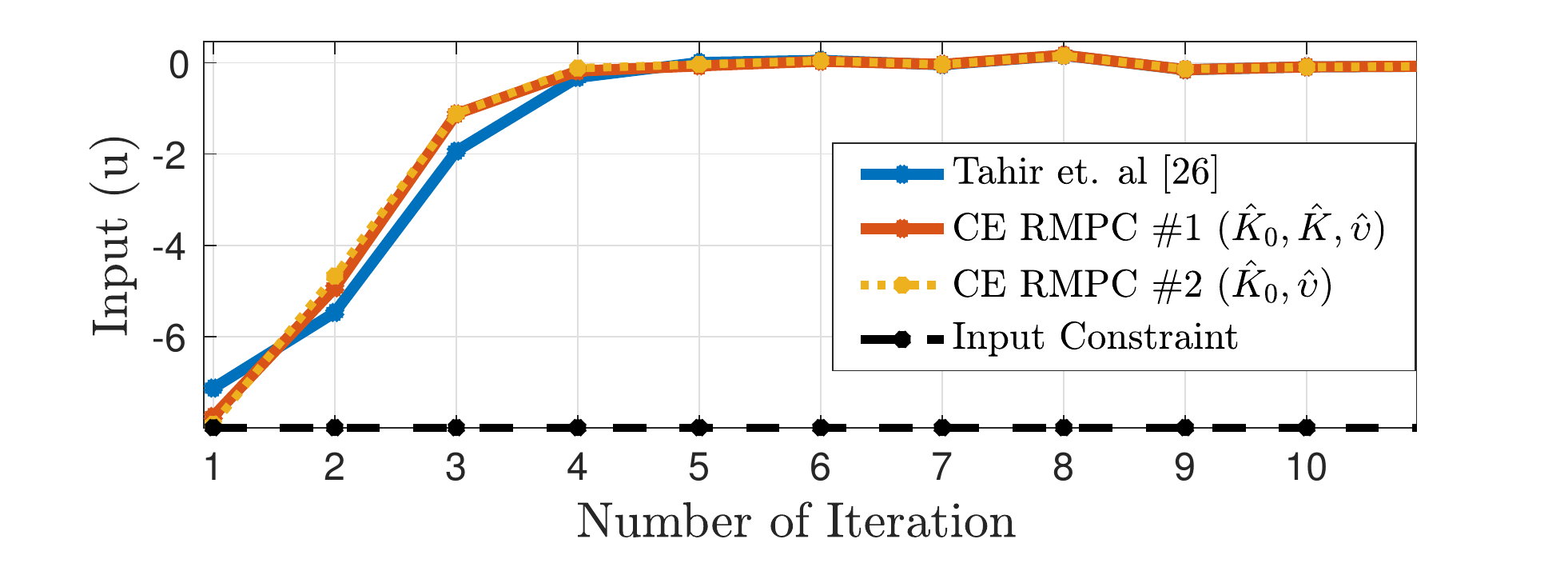}
\caption{States trajectories and control input for Example~1.}
\label{fig:OutputEx1}
\end{figure}

As shown in Fig. \ref{fig:OutputEx1},  using the algorithms from Theorem~\ref{Single_LMI} or Theorem~\ref{SngLmiThm} with fixed $\K$, robust performance has been achieved, while both states ($x_1,x_2$) converge faster to the origin compared to the results from \cite{Tahir13}. Fig.~\ref{fig:OutputEx1} also shows the control input computed by our algorithms, where it can be seen that the input reaches the constraint boundaries, in contrast with the algorithm in~\cite{Tahir13}. Therefore, our algorithms are less conservative, even for fixed $\K$, due to the novel linearization procedure (see Remark~\ref{rem:N_linearization}). The most notable outcome using our algorithms is that they required significantly lower computation burden compared to the methods from the literature. In particular, as shown in Table~\ref{table:1}, using the CE_RMPC\textit{\#1}$~\!(\K_0,\K,\hat{\upsilon})$ procedure results in the average and maximum computation cost per iteration being reduced by $94\%$ and $92\%$, respectively, as compared to the algorithm proposed in \cite{Tahir13}. Implementing CE_RMPC\textit{\#2}$~\!(\K_0,\hat{\upsilon})$ results in the average and maximum computation cost per iteration being reduced by $96\%$ and $94\%$, respectively, in comparison to the time required by the algorithm in \cite{Tahir13}. The numerical values in Table~\ref{table:1} were realized using the same computer to solve the above regulation problem and repeated 10 times for each method.  

\begin{table}[h]
\caption{Computation time per iteration for Example~1}
\label{table:1}
\centering
\begin{tabular}{ l l l }
         \hline
     Method & Mean $\pm$ Std Deviation & Maximum time \\ \hline
     Tahir et.al (2013)~\cite{Tahir13} & 23.4573 $\pm$ 2.2287 s & 29.3438 s  \\ 
     CE RMPC \textit{\#1} $(\K_0,\K,\hat{\upsilon})$ & 1.4187 $\pm$ 0.4482 s & 2.5313 s \\
     CE RMPC \textit{\#2} $(\K_0,\hat{\upsilon})$ & 0.9187 $\pm$ 0.2686 s & 1.7656 s  \\ \hline 
\end{tabular}
\end{table}

\subsection{Example 2}
The benchmark problem of control tracking of a coupled two-mass-spring system is considered \cite{Kothare96, Tahir13}. 
Discretizing the system using Euler's first order approximation for the derivative with sample time 0.1s gives a fourth order linear discrete-time state space model $x(k\!+\!1)\!=\!Ax(k)\!+\!B_uu(k)$. 
Here $A$ depends on the masses $m_1,m_2$ and the spring constant $K$.
The states $x_1,x_2$ are the mass displacements whereas $x_3,x_4$ are the respective velocities. The nominal values are $m_1\!=\!m_2\!=\!K \!=\!1$ in normalised units and the control force $u$ acts on $m_1$.

The objective is for the output ($x_2$) to track a unit step whilst satisfying the constraints $-1\!\leq\! u(k) \!\leq\! 1,~
    -\bar{x}\!\leq\! {x}(k)\! \leq \!\bar{x}\!:=\![1.5~ 1.5~ 1~ 1]^T$.
In this setup, the state is measurable, while  $K$ is 
uncertain 
within the range $0.5\!=\!K_{min}\!\leq\! K\!\leq K_{max}\!=\!10$, in appropriate units.
The cost function weights are $C_z\!=\![5I~0]^T,D_{zu}\!=\![0~I]^T$ and the prediction horizon is set as $N=6$.

\begin{figure}[h]
\centering
\includegraphics[width=\columnwidth]{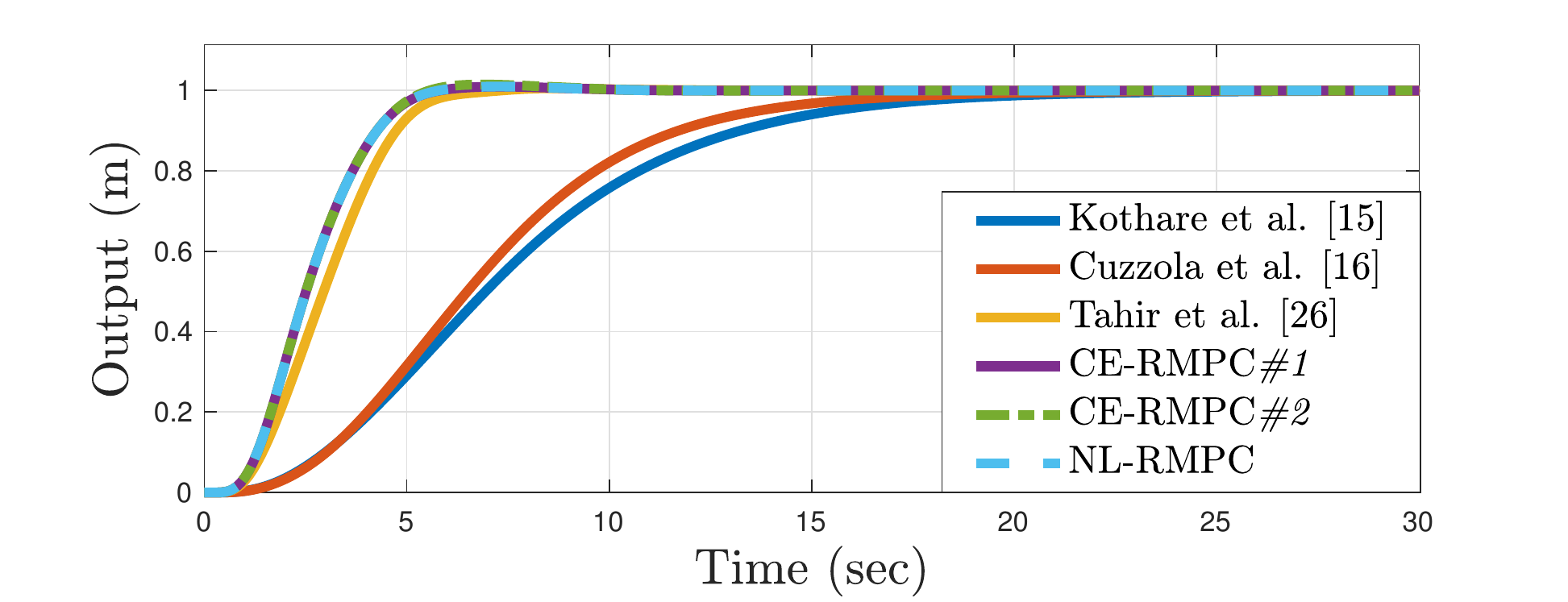}
\includegraphics[width=\columnwidth]{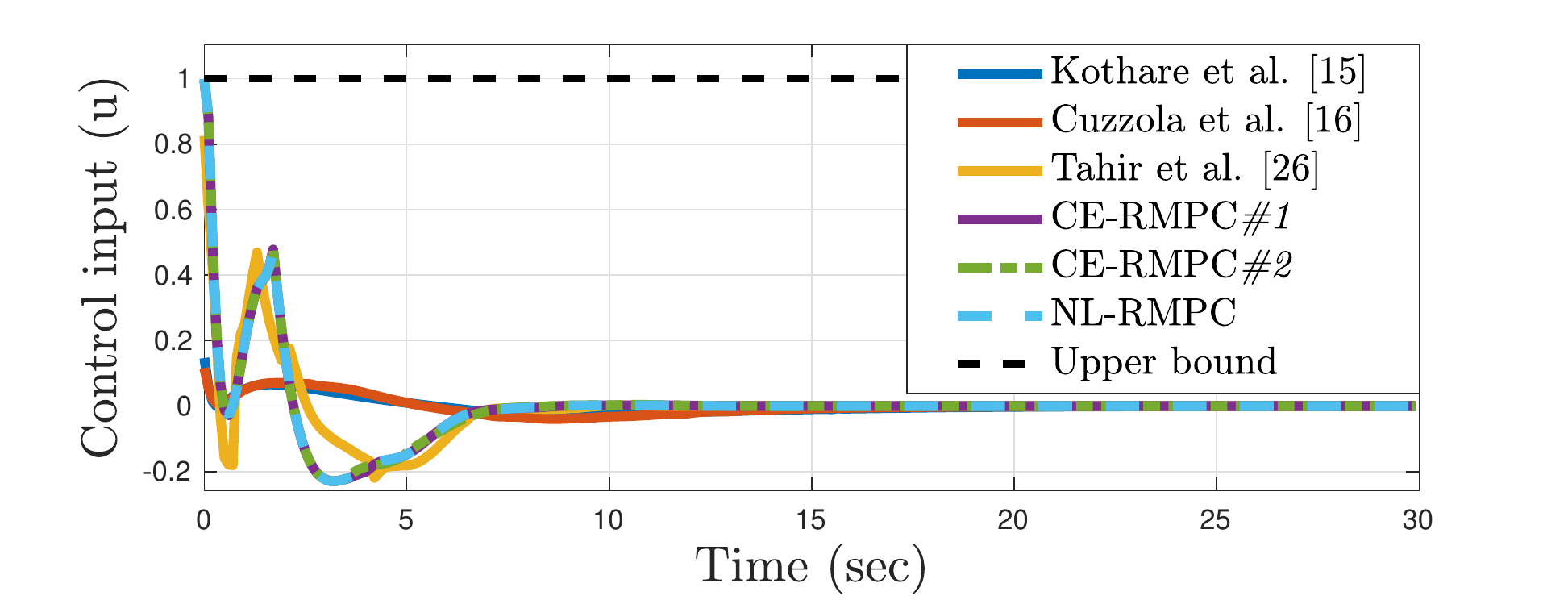}
\caption{Output and control input signals for the proposed RMPC controllers compared with algorithms from the literature for Example~2.}
\label{fig:Mass_spring_output}
\end{figure}

The output response in Fig.~\ref{fig:Mass_spring_output} shows that both proposed algorithms robustly steer the system to the reference signal. Fig.~\ref{fig:Mass_spring_output} also shows the responses of the infinite horizon methods in \cite{Kothare96, Cuzzola02} and the finite horizon method in~\cite{Tahir13}, as well as NL-RMPC$(\K_0,\K,\hat{\upsilon})$ which is described by Theorem~\ref{New_linearization} (New Linearization RMPC with multiple LMIs) . It can be seen that our approaches converge much faster than the two infinite horizon methods. Comparing with \cite{Tahir13}, it can be seen that even though all algorithms have excellent tracking properties, the proposed controllers have slightly faster responses  due to the less restrictive nature in the formulation (see Remark \ref{rem:N_linearization}). Fig.~\ref{fig:Mass_spring_output} also shows that the control input calculated by the proposed algorithms is much faster and is closer to the upper bound. Our method also gives a much smaller cost function compared with the infinite horizon methods and similar cost compared with the finite horizon method in \cite{Tahir13}.

To quantify the effect of Theorem~\ref{Single_LMI} compared with Theorem~\ref{New_linearization} with respect to feasibility domain, $K_{max}$ was increased until infeasibility is observed. CE_RMPC\textit{\#1} and CE_RMPC\textit{\#2} can obtain a solution for values up to $K_{max}=20.5$, while NL-RMPC for values up to $K_{max}=21$. Therefore we can conclude that the large computation time reduction using the suggested algorithms (CE_RMPC\textit{\#1} and CE_RMPC\textit{\#2}) comes with only a small reduction in the feasibility domain. 

Comparing the computational times in Table \ref{table:2}, it can be seen that CE_RMPC\textit{\#1} has a similar computational burden as \cite{Kothare96, Cuzzola02}, and is much faster than \cite{Tahir13} and the algorithm NL-RMPC. A significant computation time reduction can also be observed for CE_RMPC\textit{\#2}. Therefore our approach combines the fast online computational performance of the infinite horizon methods and the good performance of the finite horizon approaches. Note that in general, a larger prediction horizon increases the computation time, while stability is improved. Restricting the computational time to be similar to infinite horizon methods from the literature (on average 1 sec), the prediction horizon was set to $N\!=\!6$ to allow a fair comparison with respect to performance. Horizon length $N=7$ gives an average computational time $t=2.1845$ sec, however, the control performance was not noticeably improved.

\begin{table}[h]
\caption{Computation time per iteration for Example~2}
\label{table:2}
\centering
\begin{tabular}{ l l l }
Method & Mean $\pm$ Std Deviation & Max. time\\ \hline
Inf. horizon RMPC from~\cite{Kothare96} & 1.0788 $\pm$ 0.3321 s & 2.2813 s\\
Inf. horizon RMPC from~\cite{Cuzzola02} & 1.0679 $\pm$ 0.3002 s & 2.5625 s\\ 
RMPC from \cite{Tahir13} & 3.0672 $\pm$ 0.5137 s & 5.3750 s\\
CE RMPC \textit{\#1} & 1.0734 $\pm$ 0.2695 s & 2.0156 s \\
CE RMPC \textit{\#2} & 0.3502 $\pm$ 0.1044 s & 0.7117 s\\
NL-RMPC & 2.3547 $\pm$ 0.9762 s & 3.7656 s\\\hline 
\end{tabular}
\end{table}

\section{Conclusion}
\label{sec:Conclusion}
In this work, two algorithms are proposed to reduce the computational complexity of state-feedback RMPC for linear-time-invariant discrete-time systems, subject to structured uncertainty and bounded disturbances. 

The effectiveness of the proposed techniques is demonstrated through numerical examples taken from the literature. In particular, it has been shown that the proposed RMPC scheme can successfully calculate an optimal control signal up to $96\%$ faster than other finite horizon RMPC, while being able to steer the system quicker to a predefined reference with minimum conservativeness compared with other RMPC approaches.

Although our proposed RMPC approach is LMI-based, nevertheless, a detailed comparison with tube-based RMPC methods is currently under active research.


\bibliographystyle{IEEEtran}
\bibliography{TAC_References}

\end{document}